\documentclass[fleqn,usenatbib]{mnras}

\usepackage{newtxtext,newtxmath}

\usepackage[T1]{fontenc}

\DeclareRobustCommand{\VAN}[3]{#2}
\let\VANthebibliography\thebibliography
\def\thebibliography{\DeclareRobustCommand{\VAN}[3]{##3}\VANthebibliography}

\usepackage{graphicx}	
\usepackage{amsmath}	
\usepackage{gensymb}

\title[Stars that run away with their discs]{Double trouble: \textit{Gaia} reveals (proto)-planetary systems that may experience more than one dense star-forming environment}

\author[C. Schoettler \& R. J. Parker]{Christina Schoettler\thanks{E-mail: cschoettler1@sheffield.ac.uk}
and Richard J. Parker\thanks{Royal Society Dorothy Hodgkin Fellow}
\\
Department of Physics and Astronomy, The University of Sheffield, Hicks Building, Hounsfield Road, Sheffield S3 7RH, UK\\
}

\date{Accepted  2020 October 22. Received 2020 October 5; in original form 2020 August 28}

\pubyear{2020}

\begin{document}
\label{firstpage}
\pagerange{\pageref{firstpage}--\pageref{lastpage}}
\maketitle

\begin{abstract}
Planetary systems appear to form contemporaneously around young stars within young star-forming regions. Within these environments, the chances of survival, as well as the long-term evolution of these systems, are influenced by factors such as dynamical interactions with other stars and photoevaporation from massive stars. These interactions can also cause young stars to be ejected from their birth regions and become runaways. We present examples of such runaway stars in the vicinity of the Orion Nebula Cluster (ONC) found in \textit{Gaia} DR2 data that have retained their discs during the ejection process. Once set on their path, these runaways usually do not encounter any other dense regions that could endanger the survival of their discs or young planetary systems. However, we show that it is possible for star-disc systems, presumably ejected from one dense star-forming region, to encounter a second dense region, in our case the ONC. While the interactions of the ejected star-disc systems in the second region are unlikely to be the same as in their birth region, a second encounter will increase the risk to the disc or planetary system from malign external effects.

\end{abstract}

\begin{keywords}
stars: kinematics and dynamics -- accretion, accretion discs -- circumstellar matter -- planets and satellites: formation -- galaxies: star clusters: individual: Orion Nebula Cluster

\end{keywords}



\section{Introduction}

Many stars form grouped together with other stars in regions with stellar densities higher than are observed in the Galactic field \citep{RN25, RN59}. Gas-rich protoplanetary discs form rapidly around these stars while they are still located in high density regions \citep[e.g.][]{2001ApJ...553L.153H,2015ApJ...808L...3A,2018ApJ...869L..41A}.



These protoplanetary discs usually evolve quickly in just a few Myr into gas-poor debris discs \citep[e.g.][]{2001ApJ...553L.153H,2018MNRAS.477.5191R}, either due to the formation of planets \citep[e.g.][]{2019ApJ...884L...5S,2020ApJ...890L...9P} or truncation caused by external processes, which can also lead to the complete destruction of the discs \citep{2011ARA&A..49...67W,2015A&A...583A..66H}. Most stars lose their discs in the first 5 Myr, however, it has recently been suggested that discs can survive up to $\sim$10--12 Myr \citep{2013MNRAS.434..806B} and observations exist of even older (several tens Myr) discs around stars that are still accreting \citep[][]{2013A&A...558A.114M,2018MNRAS.476.3290M}. 


Two key factors influence the evolution and survival or destruction of the discs. One is the effect of photoevaporation due to ultraviolet (UV) radiation from nearby massive stars \citep[e.g.][]{1998ApJ...499..758J,2004ApJ...611..360A,2018MNRAS.481..452H,2019MNRAS.485.4893N,2020arXiv200607378C}. The other is the effect of dynamical interactions between the stars in regions of higher density \citep[e.g.][]{RN272,2008A&A...488..191O,RN269,2016ApJ...828...48V}. When massive stars are present in the vicinity, photoevaporation is likely to dominate \citep{RN272,2016arXiv160501773G,RN270,2018ApJ...868....1V}.

The dynamical evolution of young star-forming regions can also cause stars to be ejected \citep[e.g.][]{RN189,RN3,RN235,RN309}, possibly with their circumstellar discs intact. Not all stars that can be observed in young star-forming regions are actually born there, but can instead be visitors to the regions. It has been shown that several young visitor stars around the ONC are on a future intercept course with the cluster or have already passed through and interacted with the cluster in the past \citep{2019ApJ...884....6M,2020MNRAS.495.3104S}. If these visitors have left their birth region with an intact disc or planetary system, they can then encounter a second dense region. This means there might be young stars whose (proto)-planetary system and/or disc could suffer multiple sets of external perturbations and be subjected to ionising UV radiation more than once. 

In this letter, we search for evidence of circumstellar discs around recently ejected stars from the ONC found in \textit{Gaia} DR2 \citep{RN238}, as well as discs around future and past visitors to determine whether any (proto)-planetary systems could experience more than one dense stellar environment. In Section 2, we describe our target - the ONC and its known population of stars with circumstellar discs. Section 3 briefly describes our data analysis method. This is followed in Section 4 by the results and a brief discussion and conclusion in Section 5.

\section{Circumstellar discs in the ONC}

The Orion Nebula Cluster is a well-studied star-forming region at a distance of $\sim$400 pc \citep{RN333,RN264}. It is still very young with an estimated mean age of 2--3 Myr \citep{RN218,RN212}. Observations at different wavelengths have shown that this cluster has a population of $\sim$3500 stars \citep{1997AJ....113.1733H,1998ApJ...492..540H,2012ApJ...748...14D}. Its current average volume stellar density is approximately $4\times10{^2}$ M$_{\sun}$ pc${^{-3}}$, while its initial average volume stellar density is thought to have been much higher at 10${^3}$--10${^4}$ M$_{\sun}$ pc${^{-3}}$ \citep{RN277,RN8}. This implies that the number of dynamical encounters may have been higher in the early stages of planet formation. While higher extinction might have protected discs against radiation at these times, simulations have shown that massive stars can quickly clear out the large cavities in their immediate surroundings reducing this shielding effect \citep{2014MNRAS.442..694D}. As a result, the radiation fields experienced by discs are likely to have been much stronger than at present.

Young stars that are actively accreting from their protoplanetary discs are called Classical T Tauri stars (CTTS) (< 2 M$_{\sun}$) or Herbig Ae/Be star (2–8 M$_{\sun}$). These stars emit strongly in H$\alpha$, UV and infrared (IR) and this emission can be used to detect their discs. These emissions usually probe the inner au of the disc. Weak-lined T Tauri stars (WTTS) are thought to be more evolved than CTTS and show little or no evidence of continuing accretion from the discs in their emission \citep[e.g.][]{2015A&A...580A..26G}.



Several authors have searched for circumstellar discs around young stars in the ONC. \citet{1998AJ....116.1816H} used near-infrared (NIR) photometry combined with optical photometry and spectroscopy. They found evidence for circumstellar discs in 55--90 per cent of their sample of young stars within the mass range of 0.1--50 M$_{\sun}$. \citet{2000AJ....119.3026R} used UV excess emission in dereddened photometry to investigate the evidence for circumstellar accretion discs in the flanking fields of the ONC. They found that at least 40 per cent of the stars in their sample have a disc. \citet{2005AJ....129..363S} used H$\alpha$ profiles to find 15 new accreting member stars in the ONC. \citet{2006ApJ...646..297R} then used mid-infrared observations to study a correlation of stars with circumstellar accretion discs with their rotation period and found a clear correlation. \citet{2012AJ....144..192M} classified young stellar objects (YSO) with/without a disc via mid-infrared observations. Most recently, \citet{2019A&A...622A.149G} newly identified almost 300 young stars with discs and refined existing catalogues with new measurements.

\citet{2008ApJ...676.1109F} measured H$\alpha$ to identify accreting stars and briefly mentioned high (radial) velocity stars escaping the ONC, but found no clear disc candidates among these ejected stars. Stars can be ejected at high velocities from their birth region due to dynamical interactions or after a supernova. These stars are commonly known as runaway (RW) \citep{RN189, RN67} or walkaway (WW) stars \citep{RN136} depending on their peculiar velocity. 

The Becklin-Neugebauer (BN) object \citep{RN327} is such a fast moving, high-mass star. It is thought to have been recently ejected from the Orion region after a dynamical interaction with other ejected stars, known as Src I and Src x \citep{RN328,RN330,RN329,RN326}. Src I is of specific interest as it appears to have retained part of its disc throughout the dynamical interaction and ejection process and might even be an ejected binary system moving with a proper motion of $\sim$10 km\,s$^{-1}$ \citep{2011ApJ...728...15G,2012MNRAS.419.1390M,2020ApJ...889..178B}.

\citet{2008A&A...488..191O} investigated the effect of dynamical encounters and the impact on the circumstellar discs around higher velocity stars escaping the ONC. Their result suggested that the location of the dynamical encounters (cluster centre or outer region) and the resulting velocity of the escapers can affect the amount of disc material that remains after ejection. More recently, \citet{2019ApJ...884....6M} searched for high proper motion stars in the close vicinity of the Orion Nebula Cluster and found that seven out of their 26 candidates (including visitors and stars tracing back to other dense groups in the region) are clearly disk-bearing. The authors concluded that higher velocity are slightly more likely to be disk-less after ejection, however they do not consider the difference to be significant.

While it does not appear to be a common occurrence, these examples show that circumstellar discs can survive the ejection process and that we should in principle be able to find runaway stars with intact discs. 

\section{Method}

We use the 3D-runaway (RW) and slower walkaway (WW) results of \citet{2020MNRAS.495.3104S} as a basis to investigate if any of these stars get ejected from the ONC with an intact circumstellar disc. In addition, we also search for disc-hosting stars in the list of past visitors to the ONC from the same paper. We find secondary radial velocities (RVs) for several 2D past visitors in \citet{2018AJ....156...84K} and \citet{2015ApJ...807...27C}, where no \textit{Gaia} DR2 RVs are available. We then use this data and repeat the trace-back process described in \citet{2020MNRAS.495.3104S} to find additional past visitors to the ONC, i.e. older stars that trace back to the ONC, but have not been born there.

Following \citet{2019ApJ...884....6M}, we also search for future visitors to the ONC and use the search approach described in \citet{2020MNRAS.495.3104S}, but trace the stars' motion forwards instead of backwards in time. Given an estimated age of the ONC of $\sim$2.5 Myr \citep{RN218,RN212,2020MNRAS.495.3104S}, we search for any visitors that will travel through the ONC in the next 7.5 Myr and are currently already within 100 pc of it. This future time limit is driven by the knowledge that most young star-forming region do not live past an age of 10 Myr \citep{RN25}. We also search for secondary RVs to complement those provided in \textit{Gaia} DR2.

Using our three lists of candidates, we search through \citet{1998AJ....116.1816H}, \citet{2000AJ....119.3026R,2006ApJ...646..297R}, \citet{2005AJ....129..363S}, \citet{2008ApJ...676.1109F}, \citet{2012AJ....144..192M} and \citet{2019A&A...622A.149G} to check for the presence of a circumstellar disc around any of our past or future visitors, or ejected ONC stars.

\section{Results}

\begin{table*}
	\centering
	\caption{Stars around the ONC possibly with a circumstellar disc. Column 2+3: velocity in ONC rest frame [rf] from \citet{2020MNRAS.495.3104S}; Column 4: status identifier -- ONC ejected, future or past visitor; Column 5--7: from literature sources - $^{1}$\citet{1988AJ.....95.1744V}, $^{2}$\citet{1997AJ....113.1733H}, $^{3}$\citet{RN218}, $^{4}$\citet{2016ApJ...818...59D}: Column 8: Disc information from literature sources: [5] \citet{1998AJ....116.1816H}, [6] \citet{2000AJ....119.3026R}, [7] \citet{2006ApJ...646..297R}, [8] \citet{2008ApJ...676.1109F}, [9] \citet{2005AJ....129..363S}.}
	\label{tab:Cand_disc}
	\begin{tabular}{lcccccccc} 
		\hline
		\textit{Gaia} DR2 source-id & 2D-velocity rf & Radial velocity rf & Status identifier & Age & Mass & Spectral type & Disc information\\
		 & (km\,s$^{-1}$) & (km\,s$^{-1}$) & &  (Myr) & (M$_{\sun}$) & \\
		\hline
        3017265515291765760	&30.1 &	12.2$^{a}$	&	ONC ejected	&	0.3$^{2}$	&	2.5$^{2}$	&	K1$^{1}$ & [5]\\
        3209424108758593536 &14.1	&	-4.3$^{b}$	&	ONC ejected	&	0.5-2.5$^{3,4}$	&	0.7$^{3,4}$&	K7$^{2}$ & [5, 9]\\
        3017367151399567872  &	3.6		&	12.2$^{a}$	&	ONC ejected	& 	1.7$^{4}$	&	2.7$^{4}$	& - &	[5, 7]\\
        3209637203559481728 & 16.3 & 10.1$^{c}$ & Future visitor &  2.5$^{4}$ & 0.2$^{4}$ & - &  [6, 8]\\
        3017376325447976576 & 26.5 & -63.3$^{c}$ & Past visitor  & 18$^{4}$ & 1.4$^{4}$ & - & [6]  \\
        \hline
        \multicolumn{9}{l}{\parbox[t]{16cm}{$^{a}$\textit{Gaia} DR2, $^{b}$\citet{2015ApJ...807...27C}, $^{c}$\citet{2018AJ....156...84K}}}
	\end{tabular}
\end{table*}

Most of our fast ejected stars do not appear in any catalogue when searching for disc signatures. Nevertheless, we have identified several candidates with potential disc signatures in all three groups that might warrant further observations.

\textit{Gaia} DR2 3017265515291765760 (BD-05 1307) is the brightest 3D-RW star identified in \citet{2020MNRAS.495.3104S}. It is a very young star \citep[$\sim$0.3 Myr,][]{1997AJ....113.1733H} and is still located within the central ONC region, but will leave this region due to its high space velocity. We find this star in \citet{1998AJ....116.1816H}, who used excess emission in the IR $\Delta$(I$_{C}$-K) to identify possible circumstellar discs around young stars. These authors quoted two limiting values for the IR-excess above which a disc could be present. The first, more conservative value is ($\Delta$(I$_{C}$-K) = 0.30 mag. This RW-star has an IR-excess of ($\Delta$(I$_{C}$-K) = 0.21 mag, which falls below this higher IR-excess value. However it satisfies the lower limit of $\Delta$(I$_{C}$-K) = 0.10 mag, so a disc could be present. The IR-excess for this star is weaker than the mean and median values for stars in its local environment, suggesting it might not have originated there.

\textit{Gaia} DR2 3209424108758593536 (Brun 259) is a 3D-WW star identified in \citet{2020MNRAS.495.3104S} and also \citet{2019ApJ...884....6M}. It has an age of 0.5-2.5 Myr \citep{RN218,2016ApJ...818...59D} and is also still located within the ONC. It appears in \citet{1998AJ....116.1816H} with a value $\Delta$(I$_{C}$-K) = 0.19 mag, fulfilling their lower limit for the presence of a disc. The IR-excess for this star is weaker than the mean and median values for stars in its local environment, suggesting it might not have originated there. However, \citet{2005AJ....129..363S} stated that it is a WTTS (narrow H$\alpha$ emission), which are not expected to have much or any circumstellar material left.

\textit{Gaia} DR2 3017367151399567872 (Par 1799) is another 3D-WW candidate from \citet{2020MNRAS.495.3104S}, however its high-velocity status is driven by the RV. According to \citet{2006ApJ...646..297R}, it has indicators of a disc. These authors use a mid-IR colour index [3.6 \micro m] - [8 \micro m] > 1 mag to infer a disc and this star has a value just larger than 1 mag \citep{2006ApJ...646..297R}. This star also appears in \citet{1998AJ....116.1816H}, but has a negative $\Delta$(I$_{C}$-K), which is due to measurement errors and photometric variability.

We have also searched for disc signatures in the identified 2D-RW/WW candidates in \citet{2020MNRAS.495.3104S} and find four stars with measurements indicating the presence of discs. These stars are \textit{Gaia} DR2 3209497088842680704 and \textit{Gaia} DR2 3209498394512739968 with a small, negative UV-excess as shown in \citet{2000AJ....119.3026R}, \textit{Gaia} DR2 3014834946056441984 with a flat disc identified in \citet{2010ApJ...721..431J} and \textit{Gaia} DR2 3015714967674577024 with a transitional disc identified in \citet{2013ApJ...769..149K}. These four candidates could add to the number of ejected stars with discs, however they are missing an RV measurement and we cannot confirm their origin in the ONC yet. 

Searching through our list of future visitors to the ONC, we do not find any that have appeared in any surveys/papers searching for circumstellar discs and most of these candidates do not appear in any literature sources at all.

\citet{2019ApJ...884....6M} identified several 2D-visitors to the ONC in their paper. \textit{Gaia} DR2 3209637203559481728 (2MASS J05350504--0432334) is one such future visitor. It does not appear in our search for future visitors as we apply an astrometric quality indicator, the renormalized unit weight error (RUWE) < 1.3. This star has a much higher RUWE value indicating issues with its astrometry. It is in fact a spectroscopic binary \citep[e.g.][]{2009ApJ...697.1103T,2016ApJ...821....8K}. The proper motion errors are just $\sim$10 per cent, which is still acceptable. We can trace this star forward to the ONC in 3D using the RV from \citet{2018AJ....156...84K}, but only when considering its large distance errors (377\,$^{+57}_{-44}$ pc). It appears in \citet{2000AJ....119.3026R} with an UV-excess clearly indicative of a disc, i.e. smaller than -0.5 mag, even though it is classed as a non-accreting WTTS in \citet{2008ApJ...676.1109F}. 

There is another 2D-visitor \textit{Gaia} DR2 3017199755050720384 (V1589 Ori) identified in \citet{2019ApJ...884....6M} with a very clear disc identification \citep{2012AJ....144..192M,2019A&A...622A.149G}. Its position on the sky (within the ONC boundaries) and proper motion (moving away from the ONC) on the sky suggests a strong connection to the ONC. However based on its estimated distance of $\sim$250--350 pc \citep{RN305} and its RV \citep{2018AJ....156...84K} pointing in direction of the ONC, the star does not trace to the ONC in 3D. Based on its kinematics and position, we suggest it is not associated with the ONC (or even Orion A) at all, despite often being linked with it in literature \citep[e.g.][]{2008ApJ...676.1109F,2016ApJ...818...59D,2018AJ....156...84K}.

We have not found any stars with a disc in the past visitors list in \citet{2020MNRAS.495.3104S}. 
However, amongst the past visitors that we trace back using secondary RVs, we find one possible disc candidate. 

\textit{Gaia} DR2 3017376325447976576 (Brun 944) appeared in \citet{2000AJ....119.3026R} and showed a clear UV-excess ($\sim$-0.25 mag), which is not small enough to be directly identified as a disc candidate in that paper using their -0.5 mag upper limit. We use RV from \citet{2018AJ....156...84K} to trace back this star. It is still located inside the ONC at a distance of $\sim$390 pc \citep{RN305}, but has passed through the central, denser part already. With an estimated age of $\sim$18 Myr \citep{2016ApJ...818...59D} it cannot have been born in the ONC. 

\section{Discussion and Conclusion}

In Table \ref{tab:Cand_disc}, we present our ejected candidates with a disc. We find three recently ejected young stars from the ONC (1 RW and 2 WW stars) with some evidence of a disc, based on  IR excess observed in their emission \citep{1998AJ....116.1816H,2006ApJ...646..297R}. However, not all of these stars satisfy the more conservative excess limits stated in the above papers for a clear disc identification.

\citet{1998AJ....116.1816H} stated two IR excess limits for the identification of discs and the authors highlighted in their paper, that the stricter limit of ($\Delta$(I$_{C}$-K) = 0.30 mag might be too conservative and cause discarding of actual disc candidates. \citet{2005AJ....129..363S} stated that imposing an even higher IR excess $\Delta$(I - K) > 0.5 mag \citep[originally suggested by][]{2001AJ....122.3258R} to detect disk-bearing CTTSs can be considered a safe approach. However, they also mentioned that this can lead to missing out on a significant fraction of young stars with disks that have an IR-excess value between 0--0.5 mag and found more than one-third of the stars in this IR-excess range are CTTSs in their study \citep{2005AJ....129..363S}. The $\Delta$(I$_{C}$-K) for the two stars in Table \ref{tab:Cand_disc} are on average lower than in their local environment. This could be additional evidence that they do not originate there. However, the dynamical evolution means that the positions of individual stars are highly likely to be transient with respect to each other.

The only WW star (\textit{Gaia} DR2 3017367151399567872) that is identified as a disc candidate within the boundaries given in \citet{2006ApJ...646..297R} is not a clear-cut ejected star. It is still located within the central ONC region at a distance of $\sim$386 pc \citep{RN305} and has a 2D-velocity in the ONC reference frame that is below the escape velocity calculated by \citet{RN322}. Its RV pushes this star into the ejected star category, which could alternatively be explained as having a binary origin. Its RV has been measured by \textit{Gaia} DR2 (which we use) and \citet{2015ApJ...807...27C} and the values are consistent with each other, therefore not supporting a binary identification. This star has also featured in many studies of the ONC stellar population over the years, none of which identify a binary companion. \citet{2006A&A...458..461K} included this star as one of their targets in their search for binaries, but did not find a companion.

Just as higher RVs can be due to binary motion, proper motion can also be affected by the binary motion. The orbital motion of binaries can lead to a photocentre wobble in observations. This centroid displacement can be identified by higher RUWE values for shorter period binaries, i.e. less than the observational baseline of the survey (22 months for \textit{Gaia} DR2). For longer period binaries (several to $\sim$10 yr), this can instead lead to excess proper motion and a lower RUWE. \citep{2020MNRAS.496.1922B, 2020MNRAS.495..321P}. 

Four of the five identified star-disc candidates have a low RUWE value (< 1.3) and it is theoretically possible that the high velocities measured are at least partially due to an unknown binary companion. However, even if any of these stars were in an equal-mass binary, the average separation between the stars would not exceed 10 au (using a 10 yr period). According to estimates by \citet{2020MNRAS.496.1922B}, binaries with a semi-major axis up to 10 au should be detectable by RUWE up to a distance of 2 kpc, so these binaries would show up with a higher RUWE in addition to excess proper motion.

One of the five stars (\textit{Gaia} DR2 3209637203559481728) is a known spectroscopic binary with a high RUWE. This suggests a shorter period binary, where the measured higher proper motion is unlikely due to the binary status. The binary separation is likely to be on a similar scale to that of the disc diagnostics used ($\sim$1 au). A small separation between stars in a binary can affect any circumstellar discs present around the stars \citep[e.g.][]{2014Natur.511..567J, 2018A&A...619A.171B}

While we have not found many ejected stars that show excess emission indicative of a disc, our findings suggests that stars ejected from their birth regions due to dynamical interactions might retain some circumstellar material. We find further ejected candidates that feature in papers searching for discs, but that do not show excess emission. These stars are shown in Table \ref{tab:cand_nondisc} in Appendix \ref{non_disc} for information. While these young stars (mostly WTTS) show no indication of accretion, they could still feature harder to find debris discs or even planetary systems.

Most of the ejected stars with or without a disc are unlikely to encounter a second, dense star-forming region during their lifetime. However the location of the ONC within the Orion A molecular cloud provides several opportunities for a second encounter with such a region. In this letter, we do not trace forward the trajectories of our ejected star-disc systems that originate in the ONC. However we have searched for future visitors approaching the ONC from other regions, but have found no candidates using our more conservative search requirements.

\textit{Gaia} DR2 3209637203559481728 is a future visitor identified in \citet{2019ApJ...884....6M}, for which we find disc indicators in literature. It fulfils the strict UV excess limit of \citet{2000AJ....119.3026R}, however was classed as a WTTS by \citet{2008ApJ...676.1109F}. This young star \citep[$\sim$2.5 Myr,][]{2016ApJ...818...59D} is on approach to the ONC with a velocity of 19 km\,s$^{-1}$ (ONC reference frame) appearing to have been ejected from its birth region with a partial disc. Which part of the ONC it will encounter when passing through it in the future depends strongly on its current location. While its current position on the sky is fairly well constrained, its distance has a large margin of error (377\,$^{+57}_{-44}$ pc). It might miss the ONC completely (if the distance <$\sim$380 pc or >$\sim$415 pc), encounter only the more sparsely populated outskirts, or it may encounter the central, dense parts of the ONC. Depending on its trajectory, it might retain all of its existing circumstellar material or lose it all.

\begin{figure}
    \centering
	\includegraphics[width=\columnwidth]{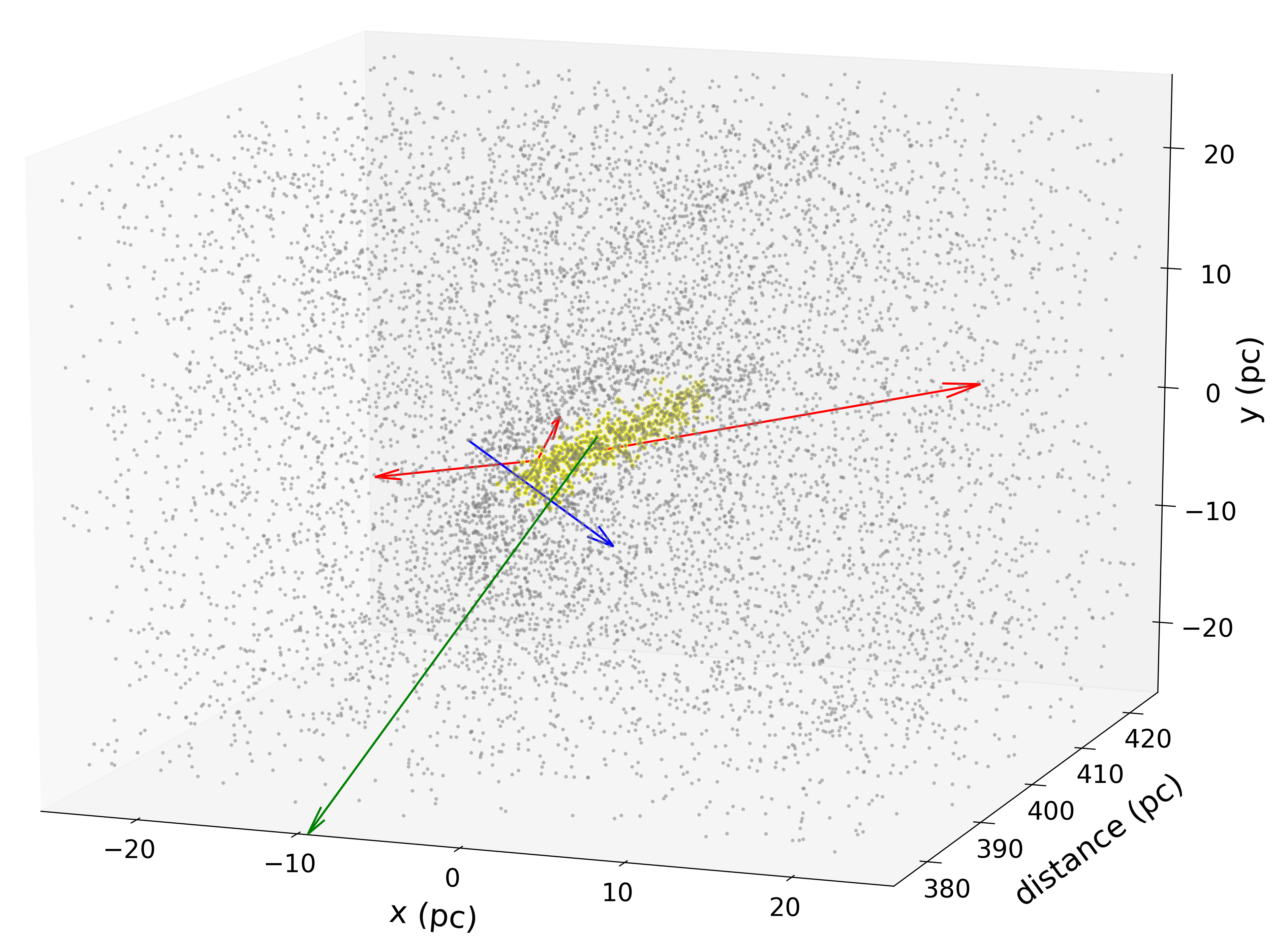}
    \caption{Location and motion of the five identified stars with possible discs. The ONC is located in the centre of the plot extending to a radius of 2.5 pc in xy-direction (plane of the sky) and 15 pc in z-direction (stars within this boundary in ``yellow'') \citep{2020MNRAS.495.3104S}. We plot fast ejected stars born in the ONC in ``red'', fast future visitors in ``blue'' and older visitors passing through the ONC in ``green''. The tail of the arrows indicates the current position of the stars, with the length of the arrows representing the scaled velocity.}
    \label{fig:discs_ONC}
\end{figure}

Finally, we find an older visitor to the ONC, that based on its estimated age \citep[$\sim$18 Myr,][]{2016ApJ...818...59D} does not originate in the ONC. While it is still located just within the ONC's boundary, it has already passed through the densest parts on its trajectory. It is the fastest of all stars identified in our search and still retains a small UV excess, but larger than the upper limit of \citet{2000AJ....119.3026R}. This type of excess emission is often used as an indicator of an accretion disc but can also point to magnetic activity in a WTTS without a disc \citep{2015A&A...581A..66V}. \citet{2000AJ....119.3026R} stated that their upper limit for UV excess \text{-0.5} mag used to distinguish disc and non-disc candidates is likely a conservative approach that might exclude stars with discs.

If this UV excess does in fact indicate the presence of a disc, then it has survived several factors that can often destroy a disc completely. Its higher peculiar velocity points to a fairly dynamical encounter leading to the ejection from its birth environment. It has then encountered a second dense region (the ONC), where it once again was subject to dynamical interactions and possibly photoevaporating radiation. Finally, its advanced age makes it less likely to still have an accretion disc. Unfortunately, this star does not appear in any other disc searches, so no final verdict on the presence/absence of a disc can be made.

While we have found a small number of possibly disc-bearing, ejected RW/WW stars, future and past visitors in our star-disc searches in literature, there is a key aspect hindering our search for RW/WW star with discs. Circumstellar discs are found predominantly around young stars, which are usually located in star-forming regions. As a consequence, observations to search for discs focus on these regions, omitting areas further away. Fig.\ref{fig:discs_ONC} indicates the positions and motions of the five identified high-velocity stars around the ONC (zoomed in to the central 50x50 pc). The figure illustrates the likely observational bias as all identified disc candidates (stars just ejected from the ONC, future visitors and older stars not born in the ONC that have visited in the past) are located in close proximity to the ONC, even though the data used cover a much larger area.

In this letter, we set out to investigate if circumstellar discs can survive the ejection from young star-forming regions using the ONC as an example. We find that there are stars at RW and WW velocities that have been ejected, which show some evidence of a disc. We also find a disc-bearing visitor from another star-forming region that is on approach to the ONC about to encounter a second higher density region. Finally, we find an older visitor that has just passed through the ONC and could possibly still have retained some of its circumstellar disc.

Whilst limited to a handful of stars, we have demonstrated that planet formation around these stars could have been hindered by external effects in more than one dense star-forming environment.

\section*{Acknowledgements}

CS acknowledges PhD funding from the 4IR STFC Centre for Doctoral Training in Data Intensive Science. RJP acknowledges support from the Royal Society in the form of a Dorothy Hodgkin Fellowship.

This work has made use of data from the European Space Agency (ESA) mission {\it Gaia} (\url{https://www.cosmos.esa.int/gaia}), processed by the {\it Gaia} Data Processing and Analysis Consortium (DPAC, \url{https://www.cosmos.esa.int/web/gaia/dpac/consortium}). Funding for the DPAC has been provided by national institutions, in particular the institutions participating in the {\it Gaia} Multilateral Agreement. 

This research has made use of the SIMBAD database, operated at CDS and the VizieR catalogue access tool, CDS, Strasbourg, France.

\section*{Data Availability Statement}
The data underlying this article were accessed from the \textit{Gaia} archive, \url{https://gea.esac.esa.int/archive/}. The derived data generated in this research will be shared on reasonable request to the corresponding author.




\bibliographystyle{mnras}
\bibliography{Main_document} 




\appendix

\section{ONC high-velocity stars without a circumstellar disc}\label{non_disc}

Table \ref{tab:cand_nondisc} shows the high-velocity stars around the ONC that were found in surveys searching for circumstellar discs but show no evidence of any disc material indicators.

\begin{table*}
	\centering
	\caption{Stars ejected from the ONC without a circumstellar disc. Column 2+3: velocity in ONC rest frame [rf] from \citet{2020MNRAS.495.3104S}; Column 4: status identifier -- ONC ejected, future or past visitor; Column 5--6: from literature sources - $^{1}$\citet{1988AJ.....95.1744V}, $^{2}$\citet{1997AJ....113.1733H}, $^{3}$\citet{RN218}, $^{4}$\citet{2012ApJ...748...14D}, $^{5}$\citet{2016ApJ...818...59D}: Column 7: Disc information from literature sources: [6] \citet{1998AJ....116.1816H}, [7] \citet{2000AJ....119.3026R}, [8] \citet{2006ApJ...646..297R}, [9] \citet{2008ApJ...676.1109F}, [10] \citet{2013ApJS..207....5F}.}
	\label{tab:cand_nondisc}
	\begin{tabular}{lcccccccc} 
		\hline
		\textit{Gaia} DR2 source-id & 2D-velocity rf & Radial velocity rf & Status identifier & Age & Mass &  Disc information source\\
		 & (km\,s$^{-1}$) & (km\,s$^{-1}$) & &  (Myr) & (M$_{\sun}$) & \\
		\hline
        3209624872711454976	&18.1 &	14.7$^{a}$	&	ONC ejected	&	0.4$^{5}$	&	0.5$^{5}$	&	 [7, 10]\\
        3017166907140904320 &17.2	&	5.3$^{b}$	&	ONC ejected	& 	1.0$^{5}$	&	0.6$^{5}$ & [7] \\	
        3017242051888552704 &16.7	&	-5.5$^{b}$	&	ONC ejected	&	1.8$^{5}$	&	0.7$^{5}$ & [7, 8]\\
        3209424108758593408  &16.4		&	8.3$^{b}$	&	ONC ejected	& 	0.5-2.5$^{3,5}$	&	1.1-2,3$^{3,5}$	& [7, 8]\\
        3017402614955763200 & 14.8 & -13.7$^{a}$ & ONC ejected &  - & - & [7]\\
        3017260022031719040 & 10.6 & 8.7$^{c}$ & ONC ejected  & 0.7–1.5$^{3,4}$ & 0.3–0.5$^{3,4}$ &  [7, 8]  \\
        3209529112120792320 & 3.0 & 12.1$^{b}$ & ONC ejected &  6.2$^{5}$ & 1.1$^{5}$ &   [7]\\
        3209531650444835840 & 13.0 & -4.2$^{a}$ & ONC ejected &  - & 3.8$^{2}$ &   [7]\\
        3017341385903759744 & 11.7 & 1.1$^{c}$ & ONC ejected &   0.7–0.9$^{3,5}$ & 0.5$^{3,5}$ &   [7]\\
        3017252600328207104 & 9.3 & -4.9$^{c}$ & ONC ejected &  0.1$^{5}$ & 0.3$^{5}$ &   [7]\\
        \hline
        \multicolumn{9}{l}{\parbox[t]{16cm}{$^{a}$\textit{Gaia} DR2, $^{b}$\citet{2015ApJ...807...27C}, $^{c}$\citet{2018AJ....156...84K}}}
	\end{tabular}
\end{table*}












\bsp	
\label{lastpage}
\end{document}